\documentclass[twocolumn]{aastex63}

\usepackage{amsfonts}
\usepackage{amsmath}
\usepackage{amssymb}
\usepackage{mathrsfs}%cursive font
\usepackage{aas_macros}
\usepackage{graphicx}
\usepackage{color}
\usepackage{float}

\def\lsim{~\rlap{$<$}{\lower 1.0ex\hbox{$\sim$}}}
\def\bsim{~\rlap{$>$}{\lower 1.0ex\hbox{$\sim$}}}

\def\yr{\ {\rm yr}}
\def\pc{\ {\rm pc}}

\def\GeV{\ {\rm GeV}}

\def\msun{\ {\rm M_\odot}}
\def\msunpppc{\ {\rm M_\odot pc^{-3}}}

\def\apjl{Astrophys.\ J.\ Lett.}

\def\aap{Astron.\ Astrophys.}

\def\apjs{Astrophys.\ J. Supp.}

\newcommand{\be}{\begin{equation}}
\newcommand{\ee}{\end{equation}}

\newcommand{\ham}{\mathscr{H}}
\newcommand{\gen}{\mathscr{W}}

\def\mathbi#1{\textbf{\em #1}}

\def\tobs{t_\text{\tiny obs}}
\def\fobs{f_\text{\tiny obs}}
\def\xobs{\xi_\text{\tiny obs}}

\def\vr{\mathbi{r}}

\def\vx{\mathbi{x}}

\definecolor{RedWine}{rgb}{0.743,0,0}
\definecolor{RoyalBlue}{rgb}{0.25,.41,.88}
\definecolor{ForestGreen}{rgb}{.13,.54,.13}
\definecolor{DeepPurple}{rgb}{.72,.18,1}

\shorttitle{Angle-Action}
\shortauthors{Desjacques et al.}

\allowdisplaybreaks

\begin{document}

\title[Axion oscillations in binary systems: angle-action surgery]{Axion oscillations in binary systems: angle-action surgery}

\author{Vincent Desjacques}
\affiliation{Physics Department, Technion -- Israel Institute of Technology, Haifa 3200003, Israel}
\affiliation{Asher Space Science Institute, Technion, Haifa 3200003, Israel}
\author{Evgeni Grishin}
\author{Yonadav Barry Ginat}
\affiliation{Physics department, Technion -- Israel Institute of Technology, Haifa 3200003, Israel}

\correspondingauthor{Vincent Desjacques}
\email{dvince@physics.technion.ac.il}

\begin{abstract}

  Scalar, tensor waves induce oscillatory perturbations in Keplerian systems that can be probed with measurements of pulsar timing residuals. In this paper, we consider the imprint of coherent oscillations produced by ultralight axion dark matter on the Roemer time delay. We use the angle-action formalism to calculate the time evolution of the observed signal and its dependence on the orbital parameters and the axion phase. We derive exact analytical expressions for arbitrary binary pulsar mass ratio and eccentricity, alleviating the need for long numerical integrations. We emphasize the similarity of the expected signal-to-noise ratio with the response of a harmonic oscillator to an external oscillatory driving. We validate our theoretical predictions with numerical simulations. Our results furnish a useful benchmark for numerical codes and analysis procedures and, hopefully, will motivate the search for such imprints in real data.
    
\end{abstract}

\keywords{binaries: general, celestial mechanics, cosmology: theory, dark matter}

%%%%%%%%%%%%%%%%%%%%%%%%%%%%%%%%%%%%%%%%%%%%%%%%%%%%%%%%%%%%%%%%%%%%%
\section{Introduction}
\label{sec:intro}
%%%%%%%%%%%%%%%%%%%%%%%%%%%%%%%%%%%%%%%%%%%%%%%%%%%%%%%%%%%%%%%%%%%%%

Gravitational radiation can resonate with binary systems and produce potentially detectable orbital perturbations \citep{rudenko:1975,mashhoon:1978,turner:1979}. When the perturbations arise from a stochastic background of gravitational waves, monitoring the randomness of the orbital elements - through the correlation functions of frequency shifts and timing residuals of pulsars, for instance - can set constraints on the amplitude of such a background \citep{mashhoon/etal:1981,mashhoon:1985,hui/etal:2011}. This effect also takes place when a binary system is excited by scalar waves \citep{annulli/etal:2018}, or embedded in a coherent background of very light bosons \citep{khmelnitsky:2013,blas/etal:2016,boskovic/etal:2018,rozner/grishin/etal:2019,Boskovic:2019qao}.

Light bosons such as QCD axions are interesting dark matter candidates \citep{preskill/etal:1983,abbott/sikivie:1983,dine/fischler:1983} because they can also resolve the strong CP problem in particle physics \citep{peccei/quinn:1977,wilczek:1978,weinberg:1978}. Ultralight axions are an extrapolation of the QCD axions (with expected masses in the range $m_a\sim 10^{-5} - 10^{-3}$ eV) to much smaller masses $m_a\sim 10^{-22} - 10^{-20}$ eV \citep{press/etal:1990,hu/etal:2000,peebles:2000}.
At low redshift, they form a Bose-Einstein condensate that oscillates coherently (unlike a stochastic background) on a timescale $\propto m_a^{-1}$ \citep[e.g.,][and references therein]{sikivie/yang:2009,marsh:2016,hui/etal:2017,niemeyer:2019,grin/etal:2019}.
Constraints on their mass $m_a$ have been set using Lyman-$\alpha$ forest measurements \citep{irsic/etal:2017,armengaud/etal:2017,kobayashi/etal:2017} and cosmic microwave background lensing \citep{hlozek/etal:2017} on megaparsec scales; dwarf spheroidals  \citep{marsh/pop:2015,gonzalez/etal:2016,marsh/niemeyer:2018,safarzadeh/spergel:2019,broadhurst/etal:2019} and ultradiffuse galaxies \citep{wasserman/etal:2019} on kiloparse scales, and galactic core observations on (sub)parsec scales \citep{desjacques/nusser:2019,bar/etal:2019,elliot/mocz:2019}.
Pulsar timing offers another avenue to probe the existence of coherent oscillations induced by ultralight scalar fields \citep{khmelnitsky:2013,blas/etal:2016,Martino2018}.
Upper limits on the amplitude of such an oscillating gravitational potential in the Milky Way halo have already been derived from pulsar timing arrays (PTAs) \citep{porayko/postnov:2014,poryako/etal:2018}. Cross-correlation of residuals from different pulsars should improve these constraints \citep{hellings/downs:1983}.

Axion coherent oscillations also resonate with binary pulsars \citep{blas/etal:2016}. While the effect is strongest near resonance, the very small width of the latter (when the coupling is purely gravitational) implies that one shall monitor instantaneous variations \citep{rozner/grishin/etal:2019} or the secular drift of orbital elements \citep{blas/etal:2019} away from resonances. 

\cite{mashhoon:1978} used Lagrange's planetary equations to develop an approximate theory of the interaction of a weak gravitational wave with a Keplerian binary. In this paper, we use angle-action variables to investigate the instantaneous variations (that is, not averaged over one orbital time) of a Keplerian system produced by an oscillating background of axion dark matter. We refer the reader to \cite{binney/tremaine:1987} for an overview of the angle-action formalism.

The paper is organized as follows. After a brief presentation of the astrophysical/cosmological context and our numerical implementation in Section \S\ref{sec:numerics}, we solve for the time evolution of the perturbed binary system using angle-action variables in Section \S\ref{sec:angleactions}. We explore the instantaneous variations of the Roemer time delay as a function of orbital parameters, etc. in Section \S\ref{sec:SNR}. We conclude in \S\ref{sec:conclusions}.

%%%%%%%%%%%%%%%%%%%%%%%%%%%%%%%%%%%%%%%%%%%%%%%%%%%%%%%%%%%%%%%%%%%%%
\section{Setup}
\label{sec:numerics}
%%%%%%%%%%%%%%%%%%%%%%%%%%%%%%%%%%%%%%%%%%%%%%%%%%%%%%%%%%%%%%%%%%%%%

We will use the numerical simulations of \cite{rozner/grishin/etal:2019} to validate our theoretical predictions. We consider a binary pulsar system with total mass $M=m_1+m_2$ and reduced mass $\mu=m_1m_2/(m_1+m_2)$.
The motion of the binary pulsar is integrated along with the perturbation induced by the coherent axion oscillations using the publicly available framework \texttt{REBOUND} \citep{ReboundMain} and the fast, adaptive, high-order integrator \texttt{IAS15} for gravitational dynamics \citep{ReboundIAS15}, accurate to machine precision over a billion orbits.
We will be interested in binaries far away from the inspiral phase so that general relativistic corrections can be neglected (see the discussion in \S\ref{sec:GR}).

In all the subsequent illustrations, we adopt the same parameters as \cite{rozner/grishin/etal:2019}, that is,
\begin{itemize}
\item a dark matter density $\rho_\text{\tiny DM}=5 \times 10^3\msunpppc$;
\item an axion mass $m_a =10^{-30}\GeV$;
\item an axion phase $\alpha=0$; and
\item a total binary mass $M=2\msun$.
\end{itemize}
The value of $\rho_\text{\tiny DM}$ is comparable to the density $\rho_c$ achieved near the hypothetical axion core (of radius $R_c\sim 1\pc$) located in the vicinity of the Milky Way halo center when the axion mass is $10^{-30}\GeV$ \citep{chavanis:2011}. In the solar neighborhood, the dark matter density is smaller by five orders of magnitude, $\rho_\text{\tiny DM}\sim 0.03\msunpppc$ \citep{salucci/etal:2010,read:2014}.
Furthermore, we conveniently define 
\be
\omega_a\equiv 2 m_a\;.
\ee
At the fundamental resonance for which $\Omega=\Omega_0=\omega_a$, the orbital frequency is $\Omega_0\simeq 3.062\times 10^{-6}$, and the semi major axis is
$a_0\simeq 0.205$ AU. 
Finally, note that the axion phase $\alpha\equiv \alpha(\vx)$ generally is a function of the spatial position $\vx$ and, thus, actually varies among binary pulsar systems.

Following \cite{rozner/grishin/etal:2019}, we will focus on the signal imprinted in the Roemer time delay, which can be extracted from measurements of the pulse times of arrival (TOAs) at the detector. The Roemer time delay is the variation of the light-travel time due to perturbations in the distance between the detector and the pulsar. In plain words, axion coherent oscillations induce a perturbation $\delta\vr(t)$ to the separation vector of the binary system at a given time $t$. Ignoring the apparent viewing geometry of the latter for simplicity, this translates into a perturbation $\Delta t_\text{\tiny TOA}(t_i)=\frac{1}{c}\big|\delta\vr(t_i)\big|$ ($c$ is the speed of light) in the pulse TOAs. The signal-to-noise ratio (S/N) for this effect can be expressed as
\be
\label{eq:RoemerSNR}
\left(\frac{S}{N}\right)^2 = \frac{1}{\sigma_\Delta^2}\sum_{i=1}^N \big(\Delta t_\text{\tiny TOA}\big)^2(t_i)\;,
\ee
where $t_i=i*\Delta$ are the times at which a TOA measurement is performed and $\sigma_\Delta$
is the error on the TOA for a pulse shape averaged over a time interval $\Delta$. In what follows, we shall adopt $\sigma_\Delta=10^{-6}$ s
for $\Delta=10$ s. The (ideal) number of measurements is $N=\tobs/\Delta$, where $\tobs$ is the total time of
observations. In practice, TOA measurements will be performed only a fraction $\fobs$ of the time. We shall hereafter assume 
$\fobs=10^{-3}$. 

Eq.(\ref{eq:RoemerSNR}) is the sole observable we shall consider here, as the Roemer delay is the largest effect in magnitude. However, the results presented in Sec. \S\ref{sec:timevol} can be used to calculate the signals imprinted in other delays \citep[see, e.g.,][for a detailed overview of timing models]{edwards/etal:2006}.

%%%%%%%%%%%%%%%%%%%%%%%%%%%%%%%%%%%%%%%%%%%%%%%%%%%%%%%%%%%%%%%%%%%%%
\section{The Axion Perturbation in Angle-Action Formalism}
\label{sec:angleactions}
%%%%%%%%%%%%%%%%%%%%%%%%%%%%%%%%%%%%%%%%%%%%%%%%%%%%%%%%%%%%%%%%%%%%%

To illustrate the power of the angle-action formalism, we will use the Delaunay variables. For simplicity, however, we will focus on the two-dimensional dynamics (justified since the angular momentum vector is conserved also in the perturbed system).
Therefore, we can restrict ourselves to the Delaunay angles $\theta_\alpha=\{\theta_b,\theta_c\}$ and actions $J_\alpha=\{J_b,J_c\}$ (they correspond to the angles $\{\theta_2,\theta_3\}$ and actions $\{J_2,J_3\}$ in \cite{binney/tremaine:1987}).
We will designate the polar variables as $q_\alpha=\{r,\vartheta\}$ and $p_\alpha=\{p_r,p_\vartheta\}$.

\subsection{Hamiltonian}

The total Hamiltonian of the system is $\ham=\ham_0+\ham_1$. The unperturbed Hamiltonian 
\be
\label{eq:H0}
\ham_0(J_c) = -\frac{\mu k^2}{2J_c^2} \;,
\ee
where $k=GM\mu$, describes the Keplerian motion. In polar coordinates, the perturbation Hamiltonian takes the form
\begin{align}
  \label{eq:Hpert}
  \ham_1 &= 2\pi G \rho_\text{DM} \mu \cos(\omega_a t + \alpha) r^2 \\
  &\equiv \mu \epsilon \Omega^2 a^2 \cos(\omega_a t + \alpha) \left(\frac{r}{a}\right)^2
  \nonumber \;.
\end{align}
Here $a$ and $\Omega$ are the semi-major axis and frequency of the binary, respectively.
The amplitude of the perturbation Hamiltonian relative to $\ham_0$ is quantified by the small parameter
\be
\epsilon = \frac{2\pi\rho_\text{\tiny DM}a^3}{M}=\epsilon_0 \left(\frac{a}{a_0}\right)^3 \;,
\ee
where $\epsilon_0$ is evaluated at the fundamental resonance. For our fiducial
parameters (see Sec.\S\ref{sec:numerics}), it is
\be
\label{eq:epsilon0}
\epsilon_0 = \frac{2\pi\rho_\text{\tiny DM}a_0^3}{M}\simeq 3.073\times 10^{-14} \;.
\ee
This justifies a perturbative treatment at first order in $\epsilon$.
The $a^3$-scaling reflects the fact that $|\ham_0|\sim \frac{1}{a}$, while 
\be
|\ham_1| \sim \epsilon \Omega^2 a^2 \sim \epsilon_0 \left(\frac{a}{a_0}\right)^3 a^{-3} a^2 \sim a^2 \;.
\ee
As expected, the effect of axion oscillations increases with the physical volume enclosed by the binary motion.

Using the definition of $\epsilon$, together with Kepler's third law $\Omega^2a^3=GM$ and the relation $a=J_c^2/\mu k$, the overall multiplicative factor in $\ham_1$ can be conveniently expressed as
\be
\frac{1}{2}\mu \epsilon \Omega^2 a^2 = \frac{1}{2} \frac{\epsilon_0}{a_0^3} \frac{J_c^4}{\mu^2 k} \;.
\ee
Furthermore, the Fourier cosine decomposition of $(r/a)^2$ on the unperturbed Keplerian orbit (justified by the smallness of $\epsilon$), for which $\theta_c$ equals the mean anomaly $\mathcal{M}$, reads
\be
\bigg(\frac{r}{a}\bigg)^2 = 1 + \frac{3}{2}e^2-\sum_{n= 1}^\infty \frac{4}{n^2} J_n(ne) \cos(n\theta_c) \;,
\ee
Here and henceforth, $0\leq e<1$ will denote the eccentricity.
As a result, the perturbation Hamiltonian in the angle-action variables can be recast into the form
\begin{widetext}
\be
\label{eq:Hpert1}
\ham_1(\theta_c,J_b,J_c;t) = \frac{1}{2} \frac{\epsilon_0}{a_0^3} \frac{J_c^4}{\mu^2 k}
\bigg\{\cos(\omega_a t + \alpha) \left(5-3\frac{J_b^2}{J_c^2}\right) \\
- \sum_{n= 1}^\infty\frac{4}{n^2} J_n(ne) \Big[\cos\!\big(\omega_at + n\theta_c + \alpha\big)
  +\cos\!\big(\omega_a t-n\theta_c+\alpha\big)\Big]\bigg\} 
\ee
\end{widetext}
where it is understood that $e\equiv \sqrt{1-J_b^2/J_c^2}$.
Note also that $\frac{\epsilon_0}{a_0^3}\frac{J_c^3}{\mu^2 k}=\epsilon\Omega$. 

Eq.(\ref{eq:Hpert1}) will be useful for the computation of the time evolution of the perturbations (presented in Sec. \S\ref{sec:timevol}).

\subsection{Perturbed Displacement}
\label{sec:displacement}

In the variables $(\theta_b,\theta_c,J_b,J_c)$, the separation vector $\vr(t)$ takes the general form
\be
\vr(t) = \vr\big(\theta_b(t),\theta_c(t),J_b(t),J_c(t)\big) \;.
\ee
The time dependence of the angle-action variables is governed by Hamilton equations:
\begin{gather}
\dot{\theta}_\alpha = + \frac{\partial\ham}{\partial J_\alpha} \;,\qquad
\dot{J}_\alpha = - \frac{\partial\ham}{\partial \theta_\alpha} \;, \\
\ham = \ham_0(J_c) + \ham_1(\theta_c,J_b,J_c;t) \nonumber \;.
\end{gather}
The unperturbed solution given by $\ham_0$ is
\be
\vr_0(t)=\vr\big(\theta_b^0(t),\theta_c^0(t),J_b^0(t),J_c^0(t)\big) 
\ee
with
\begin{gather}
\theta_b^0(t) = 0 \;,\qquad \theta_c^0(t) = \mathcal{M}(t) \;, \\
J_b^0 = J_c^0\sqrt{1-e^2} \;,\qquad J_c^0 = \sqrt{k \mu a}
\nonumber \;.
\end{gather}
It will be convenient to parameterize the unperturbed orbit in terms of the eccentric anomaly $\xi$, and express the mean anomaly $\mathcal{M}$ as $\mathcal{M}(\xi)=\xi-e\sin\xi$.

Combining the previous expressions, the displacement vector reads
\begin{align}
\delta\vr(t) &\equiv \vr(t) - \vr_0(t) \\
&\approx (\cos\vartheta,\sin\vartheta)\,\delta r + r (-\sin\vartheta,\cos\vartheta)\,\delta\vartheta 
\nonumber
\end{align}
at first  order in the small perturbation $\epsilon\ll 1$, with
\begin{align}
  \label{eq:ddisplacement}
  \delta r &= \frac{\partial r}{\partial\theta_\alpha}\bigg\lvert_0\Delta\theta_\alpha+\frac{\partial r}{\partial J_\alpha}\bigg\lvert_0\Delta J_\alpha \\
  \delta\vartheta &= \frac{\partial\vartheta}{\partial\theta_\alpha}\bigg\lvert_0\Delta\theta_\alpha +\frac{\partial\vartheta}{\partial J_\alpha}\bigg\lvert_0\Delta J_\alpha
  \nonumber \;.
\end{align}
These relations follow from writing the position vector in polar coordinates, $\vr=r(\cos\vartheta,\sin\vartheta)$.
Since the perturbations $\Delta\theta_\alpha(t)$ and $\Delta J_\alpha(t)$ are first order in $\epsilon$, the partial derivatives of the polar
coordinates are computed on the unperturbed orbit.
Furthermore, the Einstein summation convention is implied here and throughout this paper.

To calculate the derivatives of the polar coordinates with respect to the angle-action variables, we start from the function generating the canonical transformation to the Delaunay variables. The details and results of the computation can be found in Appendix \S\ref{sec:generating}.

To conclude this Section, one could in principle transform to a new set of angle-action coordinates constructed such that the perturbed Hamiltonian depends on the new action variables solely (up to first order in $\epsilon$). This standard procedure is briefly reviewed in, e.g., \cite{annulli/etal:2018}. However, it does not give any practical advantage in the computation of the signal we are aiming at. 
Therefore, we have not implemented it here.

\subsection{Time Evolution}
\label{sec:timevol}

Next, we compute the perturbations $\Delta\theta_\alpha(t)$ and $\Delta J_\alpha(t)$ to the angle-action variables from the perturbation Hamiltonian Eq.(\ref{eq:Hpert}).
We require that the perturbed and unperturbed orbits coincide initially (that is, at the beginning of the observational period), so that $\Delta\theta_\alpha$ and $\Delta J_\alpha$ vanish at $t=0$. We will denote the initial eccentric and mean anomaly as $\xi_0$ and $\mathcal{M}_0$, respectively. On the unperturbed trajectory, we thus have $\theta_c^0(t) = \mathcal{M}(t) = \Omega t + \mathcal{M}_0$, with $\mathcal{M}_0=\xi_0-e\sin\xi_0$. The offset between the periapsis passages and the peaks of the axion oscillatory forcing evolves with time depending on the initial conditions ($\alpha$,$\mathcal{M}_0$) and the frequencies $(\omega_a,\Omega)$ unless one sits at a resonance (in which case only $\alpha$ and $\mathcal{M}_0$ matter).

For the angles, Hamilton equations give
\begin{widetext}
\begin{align}
  \Delta\theta_b(t) &= \int_0^t\!dt'\,\frac{\partial\ham_1}{\partial J_b} \label{eq:DeltaTbt} \\
  &= -3\epsilon\left(\frac{\Omega}{\omega_a}\right)\sqrt{1-e^2}\;\Big(\sin(\omega_a t+\alpha)-\sin\alpha\Big) 
  +2\epsilon\left(\frac{\Omega}{\omega_a}\right)\frac{\sqrt{1-e^2}}{e}
  \sum_{n=1}^\infty \frac{J_n'(ne)}{n}\mathcal{S}_{n1}^{(+)}\!(\omega_a,\Omega,\alpha,\mathcal{M}_0;t) \nonumber \\
  \Delta\theta_c(t) &= \int_0^t\!dt'\,\bigg(-3 \frac{\Omega}{J_c}\Delta J_c+\frac{\partial\ham_1}{\partial J_c}\bigg) \label{eq:DeltaTct} \\
  &= -6 \epsilon \left(\frac{\Omega}{\omega_a}\right)\, \sum_{n=1}^\infty\frac{J_n(ne)}{n} \bigg[\left(\frac{\Omega}{\omega_a}\right)
      \mathcal{S}_{n2}^{(-)}\!(\omega_a,\Omega,\alpha,\mathcal{M}_0;t)-\Omega t\,\mathcal{A}_n\!\left(\frac{\Omega}{\omega_a},\alpha,\mathcal{M}_0\right) \bigg]
  +\epsilon\left(\frac{\Omega}{\omega_a}\right)\big(7+3e^2\big)\nonumber \\
  &\qquad \times \Big(\sin(\omega_a t+\alpha)-\sin\alpha\Big)
  -2\epsilon\left(\frac{\Omega}{\omega_a}\right)
  \sum_{n= 1}^\infty\left[4\frac{J_n(ne)}{n^2}+\left(\frac{1-e^2}{e}\right)\frac{J_n'(ne)}{n}\right]
  \mathcal{S}_{n1}^{(+)}\!(\omega_a,\Omega,\alpha,\mathcal{M}_0;t) \nonumber
\end{align}
whereas for the actions we have
\be
  \label{eq:DeltaJbt} 
  \Delta J_b(t) = -\int_0^t\!dt'\,\frac{\partial\ham_1}{\partial\theta_b} 
  = 0\;,
\ee
which expresses the conservation of angular momentum, and 
\be
  \label{eq:DeltaJct} 
  \Delta J_c(t) = -\int_0^t\!dt'\,\frac{\partial\ham_1}{\partial\theta_c} =2\epsilon\left(\frac{\Omega}{\omega_a}\right) J_c\sum_{n= 1}^\infty\frac{J_n(ne)}{n}\,\mathcal{C}_{n1}^{(-)}\!(\omega_a,\Omega,\alpha,\mathcal{M}_0;t) \;.
\ee
To derive all these expressions, we have substituted the unperturbed solution $\theta_c=\theta_c^0(t)$, and taken advantage of the
relations $J_c=\mu \Omega a^2$ and $\Omega = \mu k^2/J_c^3$ to simplify them further.

For shorthand convenience, we have also introduced the functions $\mathcal{S}^{(\pm)}_{nq}\!(\omega_a,\Omega,\alpha,\mathcal{M}_0;t)$ and $\mathcal{C}^{(\pm)}_{nq}\!(\omega_a,\Omega,\alpha,\mathcal{M}_0;t)$
defined as
\begin{align}
  \mathcal{S}_{nq}^{(\pm)} &=
  \frac{\sin\!\big(\omega_at +n\Omega t+\alpha+n\mathcal{M}_0\big)-\sin(\alpha+n\mathcal{M}_0)}{\left(1+n\frac{\Omega}{\omega_a}\right)^q}\pm
  \frac{\sin\!\big(\omega_at -n\Omega t+\alpha-n\mathcal{M}_0\big)-\sin(\alpha-n\mathcal{M}_0)}{\left(1-n\frac{\Omega}{\omega_a}\right)^q} \;, \\
  \mathcal{C}_{nq}^{(\pm)} &=
  \frac{\cos\!\big(\omega_a t+n\Omega t+\alpha+n\mathcal{M}_0\big)-\cos(\alpha+n\mathcal{M}_0)}{\left(1+n\frac{\Omega}{\omega_a}\right)^q}\pm
  \frac{\cos\!\big(\omega_a t-n\Omega t+\alpha-n\mathcal{M}_0\big)-\cos(\alpha-n\mathcal{M}_0)}{\left(1-n\frac{\Omega}{\omega_a}\right)^q}
  \nonumber \;.
\end{align}
\end{widetext}
The $(\pm)$ determines their parity under the transformation $n\to -n$.
Furthermore, both $\mathcal{S}^{(\pm)}_{nq}$ and $\mathcal{C}^{(\pm)}_{nq}$ vanish at the initial time $t=0$.

%------------------------------------------------------------
\begin{figure*}
  \centering
  \includegraphics[width=14cm]{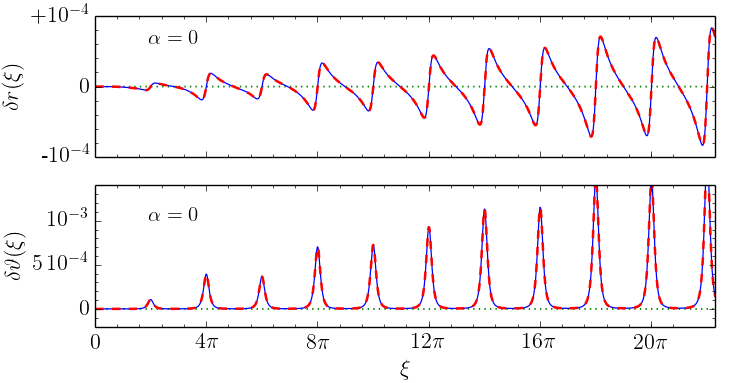}
  \includegraphics[width=14cm]{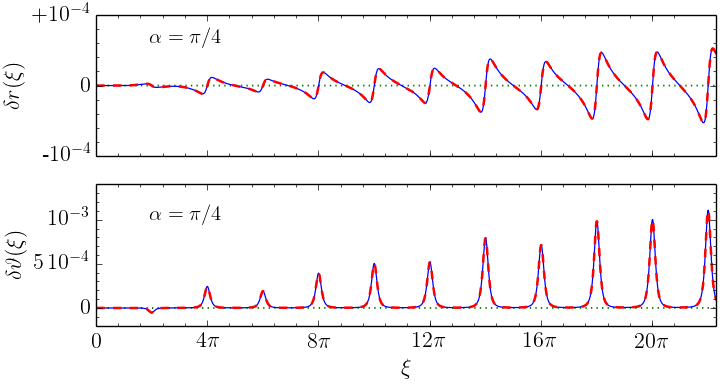}
  \caption{Time evolution of the coordinate perturbations $\delta r(\xi)$ and $\delta\vartheta(\xi)$ as a function of the eccentric anomaly $\xi$
    (see text for details). The solid (blue) curve is the result of a direct numerical integration in polar coordinates, while the dashed (red)
    curve shows the theoretical prediction (at first order in $\epsilon$). The phase is $\alpha=0$ (top panels) and $\alpha=\pi/4$ (bottom panels).
    The other, common parameter values are $\epsilon_0=10^{-7}$, $\frac{\Omega}{\omega_a}\approx 1.713$, and $e\approx 0.866$. Note that $\delta r$
    is plotted in units of the semi-major axis $a$.}
  \label{fig:Orbit}
\end{figure*}
%------------------------------------------------------------

For $\Delta \theta_c$, the expression is somewhat more involved because one needs to consider two variations:
\be
\Delta \dot{\theta}_c = \Delta\left(\frac{\partial\ham_0}{\partial J_c}\right) + \frac{\partial\ham_1}{\partial J_c}
= \frac{\partial\Omega}{\partial J_c}\Delta J_c + \frac{\partial\ham_1}{\partial J_c}\;.
\ee
The first term on the right-hand side is the perturbation to the Keplerian frequency. 
As a result, the angle $\theta_c$ evolves faster (or slower) relative to the unperturbed case. We have
\be
  \frac{\partial \Omega}{\partial J_c}{\Delta J_c}= -3 \frac{\Omega}{J_c}\Delta J_c \nonumber \;.
\ee
This effect vanishes at first order for a perfectly circular orbit ($e=0$) because the unperturbed orbit sits at the bottom of the effective one-body (radial)
potential. As a consequence, any variation in the frequency - or energy - must be second order for the circular case.
Using Eq.(\ref{eq:DeltaJct}) and integrating over time, this becomes
\begin{widetext}
\be
\label{eq:ChangeOrbitalFrequency}
-3 \frac{\Omega}{J_c}\int_0^t\!dt'\,\Delta J_c(t') =-6 \epsilon \left(\frac{\Omega}{\omega_a}\right)\,
\sum_{n= 1}^\infty\frac{J_n(ne)}{n}
\bigg[\left(\frac{\Omega}{\omega_a}\right)\mathcal{S}_{n2}^{(-)}(\omega_a,\Omega,\mathcal{M}_0,\alpha;t) \\
  -\Omega t\,\mathcal{A}_{n}\!\left(\frac{\Omega}{\omega_a},\alpha,\mathcal{M}_0,\right)\bigg] \;,
\ee
\end{widetext}
in which the time-independent function $\mathcal{A}_n(x,\alpha,\mathcal{M}_0)$ is
\be
\mathcal{A}_n = \frac{\cos(\alpha-n\mathcal{M}_0)}{\big(1-n x\big)}-\frac{\cos(\alpha+n\mathcal{M}_0)}{\big(1+n x\big)} \;.
\ee
This is the first term on the right-hand side of Eq.(\ref{eq:DeltaTct}). Observe that, in the limit $\frac{\Omega}{\omega_a}\to 0$, the term in square brackets scales like $\Omega^2$, and consequently this effect becomes subdominant when $a\gg a_0$.

To validate our analytical results, we numerically evolved the perturbed and unperturbed system in polar coordinates and
extracted the time evolution of $\delta r$ and $\delta\vartheta$, which we  compared to our theoretical prediction obtained upon combining
Eqs.~(\ref{eq:DeltaTbt}) -- (\ref{eq:DeltaJct}) with the partial derivatives Equations~(\ref{eq:dradial}) and~(\ref{eq:dtheta}).
In practice, we truncated the series expansion of $\ham_1$ at the 20th harmonic. 
The results are shown in Fig.\ref{fig:Orbit} as a function of the eccentric anomaly $\xi$. The initial conditions were set at pericenter passage ($\mathcal{M}_0=\xi_0=0$).
They assume a highly eccentric orbit with $e\approx 0.866$. As a consequence, the change in orbital frequency is the dominant effect. This translates into a fairly large perturbation in the polar angle $\delta\vartheta$ (relative to $\delta r$) owing to the term $\frac{\partial\vartheta}{\partial\theta_c}\Delta\theta_c$, which peaks at pericenter passage.
Note that $\delta r$ is shown in units of the semi-major axis $a$.

We emphasize that our calculation is valid everywhere except in small neighborhoods of size $\sqrt{\epsilon}$ centered on the resonances. 
The near-resonance case is thoroughly discussed in \cite{blas/etal:2016} and \cite{rozner/grishin/etal:2019}.
Note also that the angle perturbations given by Equations (\ref{eq:DeltaTbt}) - (\ref{eq:DeltaTct}) do not depend on the reduced mass $\mu$ of the system, while Eq.(\ref{eq:DeltaJct}) does through the multiplicative factor of $J_c$. However, the latter cancels out in $\delta r$ and $\delta\vartheta$. Hence, the perturbation $\delta\vr$ to the separation vector truly is independent of the reduced mass, as requested by the equivalence principle.

%%%%%%%%%%%%%%%%%%%%%%%%%%%%%%%%%%%%%%%%%%%%%%%%%%%%%%%%%%%%%%%%%%%%%
\section{Signal-to-noise Ratio for the Roemer Delay}
\label{sec:SNR}
%%%%%%%%%%%%%%%%%%%%%%%%%%%%%%%%%%%%%%%%%%%%%%%%%%%%%%%%%%%%%%%%%%%%%

Having solved the equations of motion to first order in $\epsilon$, we will now concentrate on the variations in Roemer time delay and the corresponding S/N as quantified in Sec. \S\ref{sec:numerics}.

\subsection{General Expression}

Taking into account a duty cycle of $\fobs$ as advocated above, the S/N Eq.(\ref{eq:RoemerSNR}) for the perturbation to the Roemer time delay can be expressed as
\begin{align}
  \label{eq:RoemerSNRt}
  \left(\frac{S}{N}\right)^2 &= \frac{\fobs}{\sigma_\Delta^2\Delta} \sum_{i=1}^N \Delta\cdot \big(\Delta t_\text{\tiny TOA}\big)^2(t_i) \\
  &= \frac{\fobs}{\sigma_\Delta^2\Delta}\int_0^{\tobs}\!dt\,\frac{\big|\delta\vr\big|^2}{c^2} \bigg. \nonumber \\
  &= \epsilon^2a^2\left(\frac{\Omega}{\omega_a}\right)^2\left(\frac{\fobs\tobs}{\sigma_\Delta^2c^2\Delta}\right)
  \left\{\frac{1}{\tobs}\int_0^{\tobs}\!dt\,\big|\delta\tilde\vr\big|^2\!(t)\right\} \nonumber
\end{align}
where, for convenience, we have introduced a dimensionless displacement $\delta\tilde\vr(t)$ defined through the relation 
\be
\delta\vr(t) \equiv \epsilon a \left(\frac{\Omega}{\omega_a}\right)\, \delta\tilde\vr(t) \;.
\ee
The last two equalities in Eq.(\ref{eq:RoemerSNRt}) assume that the sum over discrete times can be traded for an integral, which is a good approximation when the number of measurements is large.

To calculate the S/N, an expression for $|\delta\vr|^2=\delta r^2 + r^2 \delta\vartheta^2$, where $\delta r$ and $\delta\vartheta$ are given by Eq.~(\ref{eq:ddisplacement}), is required.
For this purpose, it is convenient to use the eccentric anomaly $\xi$ as time variable and recast Eq.(\ref{eq:RoemerSNRt}) into
\begin{align}
  \label{eq:RoemerSNRxi}
  \left(\frac{S}{N}\right)^2 &= \left(\frac{\epsilon\,\Omega\, a}{\omega_a}\right)^2\left(\frac{\fobs\tobs}{\sigma_\Delta^2c^2\Delta}\right)\\
  &\quad \times
  \left\{\frac{1}{\Omega\tobs}\int_{\xi_0}^{\xobs+\xi_0}\!\!d\xi\,\big(1-e\cos\xi\big)\,\big|\delta\tilde\vr\big|^2\!(\xi)\right\} \nonumber \;,
\end{align}
where $\xobs$ denotes the amount of eccentric anomaly elapsed during the observational run.
The upper limit $\xobs=\xobs(\tobs)$ of the integral is determined from $\Omega\tobs = \mathcal{M}(\xobs)-\mathcal{M}(\xi_0)$.

To get insight into the dependence of the S/N on the axion mass and the orbital parameters, consider the limit $\Omega\gg \omega_a$. In this regime, the perturbation produced by the axion coherent oscillations can be treated as time independent. Therefore, Eq.(\ref{eq:Hpert}) shows that they yield a force of amplitude $\epsilon \Omega r\sim \epsilon\Omega a$ per unit mass.
This implies that the variation $\delta\dot a$ is
\be
\delta\dot a = \int_0^T\!dt\,\delta\ddot r \sim \epsilon \Omega^2 a \int_0^T\!dt \sim \epsilon \Omega a 
\ee
since the orbital period is $T=2\pi/\Omega$. For a total observational time $\tobs$, the change $\delta a$ thus is
\be
\label{eq:deltaascaling}
\delta a \sim \int_0^{\tobs}\!dt\,\delta\dot a \sim \epsilon \Omega a\, \tobs \sim a^{5/2}\,\tobs \;.
\ee
Since the overall amplitude of the signal-to-noise ratio is proportional to
\be
\label{eq:gSNRscaling}
\epsilon\cdot \Omega\cdot a \propto a^3 \cdot a^{-3/2} \cdot a \propto a^{5/2} \;,
\ee
Eq.(\ref{eq:deltaascaling}) suggests that the SNR should behave like $a^{5/2}$ in the limit $a\to 0$ or, equivalently, the curly brackets in Eq.(\ref{eq:RoemerSNRxi}) asymptotes to
a constant in the same limit. We will see that this is indeed the case.

\subsection{Development at small eccentricities}

Emission of gravitational waves will eventually circularize the orbit of binary pulsar systems, so that $e\approx 0$ is a very good approximation in the late
stage of the coalescence phase \citep{peters:1964}. Therefore, it is instructive to develop the previous results for low eccentricities. As we shall see now, the S/N can be cast into a simple expression in the limit $e\to 0$. Details can be found in Appendix \S\ref{sec:smalleccentricities}

The square $|\delta\tilde\vr|^2$ of the normalized perturbed displacement can eventually be expressed as
\begin{widetext}
\be
\label{eq:tildedeltare0}
  |\delta\tilde\vr|^2 = 
  \bigg[\sin\xi\,\mathcal{S}_{11}^{(+)}+\cos\xi\,\mathcal{C}_{11}^{(-)}\bigg]^2
  + 4\bigg[2\Big(\sin(\omega_a t+\alpha)-\sin\alpha\Big)+\sin\xi\,\mathcal{C}_{11}^{(-)}-\cos\xi\,\mathcal{S}_{11}^{(+)}\bigg]^2
  +\mathcal{O}(e) \;.
\ee
\end{widetext}
Integrating this expression from $t=0$ until $t=\tobs$ returns terms linear in $\tobs$ along with a transient contribution that vanishes in the limit $\tobs\to\infty$.
We shall mainly focus on the former in the following discussion since it solely survives for large $\xobs$, which is the experimental setup
considered here. However, we will also show the full result for the sake of comparison with the data.

%------------------------------------------------------------
\begin{figure}
  \centering
  \includegraphics[width=9cm]{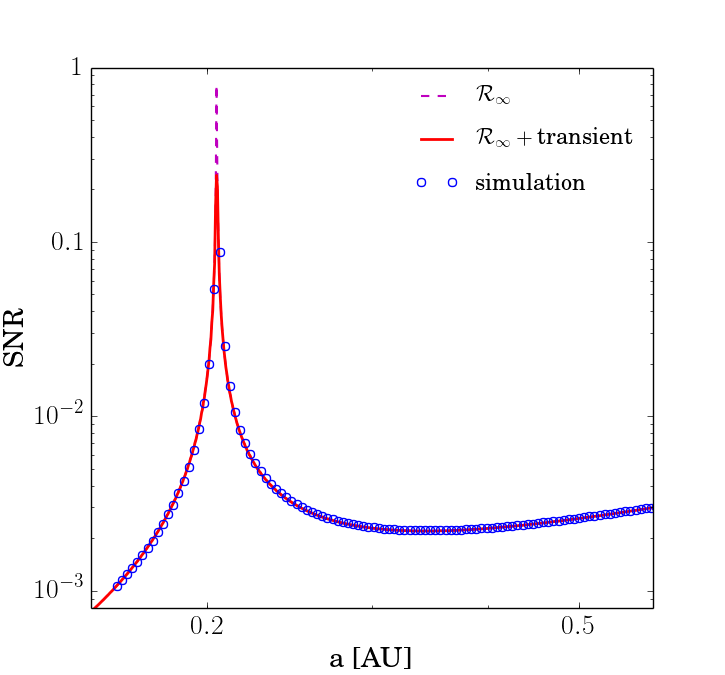}
  \caption{S/N for a detection of the Roemer time delay in the case of a circular orbit ($e=0$).
    The blue data points are the simulations, the solid (red) curve represents the long-time asymptotics prediction,
    Eq.(\ref{eq:Responsezeroe}), while the dashed (magenta) curve shows the full result, including the transient contribution Eq.(\ref{eq:Transientzeroe}). For this run, we have artificially increased $\epsilon_0$ by a factor of $10^5$ in the simulations in order to mitigate the numerical noise. We have rescaled the final results so that they correspond to the fiducial model outlined in Sec. \S\ref{sec:numerics}.}
  \label{fig:SNRe0}
\end{figure}
%------------------------------------------------------------

\subsection{The Case $e=0$}
\label{sec:zeroe}

For a circular orbit, the value of the initial mean anomaly $\mathcal{M}_0$ and the phase $\alpha$ of the axion field are irrelevant to the S/N. Therefore, we can choose $\mathcal{M}_0=\alpha=0$ without any restriction.
With this simplification, a straightforward calculation shows that 
\be
  \frac{1}{\xobs}\int_0^{\xi_{\rm obs}}\!\!d\xi\,\big|\delta\tilde\vr\big|^2\!(\xi) = \mathcal{R}_\infty\!\!\left(\frac{\Omega}{\omega_a}\right)
  + \mbox{transient} \;.
\ee
The transient contribution is of the form
\be
\label{eq:Transientzeroe}
\sum_i c_i \frac{\sin(\varpi_i\xobs)}{\varpi_i\xobs}\;,
\ee
where the various amplitudes $c_i$ and frequencies $\varpi_i$ (loosely labelled with an index $i$) are functions of the frequencies $m_a$ and $\Omega$. Its explicit expression -- which is subdominant in the long-time asymptotic limits -- is too long to be given here.
The response function $\mathcal{R}_\infty(x)$ -- which dominates in the long-time asymptotic limit -- takes the form
\be
\label{eq:Responsezeroe}
\mathcal{R}_\infty\!(x) = 8 x^2 \frac{\left(\frac{3}{2}+x^2\right)}{\left(1-x^2\right)^2} \;.
\ee
It diverges at the fundamental resonance $x=1$, and its asymptotic behavior as the argument tends toward zero or infinity is
\be
\mathcal{R}_\infty\!(x) = \left\lbrace\begin{array}{ll} \Big. 12 x^2 + \mathcal{O}(x^4) & (x\to 0) \\
\Big. 8+28x^{-2}+\mathcal{O}(x^{-4}) & (x\to\infty)\end{array} \right. \;.
\ee
Unsurprisingly, $\mathcal{R}_\infty\!(x)$ is analogous to the response function (or transfer function) of a standard driven harmonic oscillator of frequency $\Omega$, in which the axion oscillations
$\propto \sin(\Omega t/x+ \alpha)$ play the role of the external driving force (this can also be seen upon writing the orbit equation for $u(\vartheta)$, where $u\equiv 1/r$).
In the limit $x\to \infty$ (slow driving), this force can be treated as constant and, therefore $\mathcal{R}_\infty(x)\to$ const. since the displacement is independent of frequency.
In the limit $x\to 0$ (fast driving), the constant (i.e. frequency-independent) terms in Eq.(\ref{eq:tildedeltare0}) cancel out so that the response function is proportional to
$x^2\sim (\omega_a)^{-2}$. All this remains true when $e>0$ (see Sec.~\S\ref{sec:nonzeroe}).

Substituting the response function into Eq.(\ref{eq:RoemerSNRxi}), the long-time signal-to-noise reads
\be
\label{eq:SNRe0asym}
  \left(\frac{S}{N}\right)^2 = \left(\frac{\epsilon\, \Omega\, a}{\omega_a}\right)^2 \frac{f_\text{obs} t_\text{obs}}{\sigma_\Delta^2c^2\Delta}\,\,
  \mathcal{R}_\infty\!\!\left(\frac{\Omega}{\omega_a}\right) \;.
\ee
Using Eq.(\ref{eq:gSNRscaling}), the SNR expressed as a function of the semi-major axis $a$ scales like
\be
\left(\frac{S}{N}\right) \propto \left\lbrace\begin{array}{ll} a^{5/2} & (a\ll a_0) \\ a & (a\gg a_0)  \end{array}\right.
\ee
The $a^{5/2}$ behaviour in the regime $a\ll a_0$ reflects the scaling $\mathcal{R}_\infty(x)\to$ const. of the response function for large orbital frequencies $x\gg 1$.

%------------------------------------------------------------
\begin{figure}
  \centering
  \includegraphics[width=9cm]{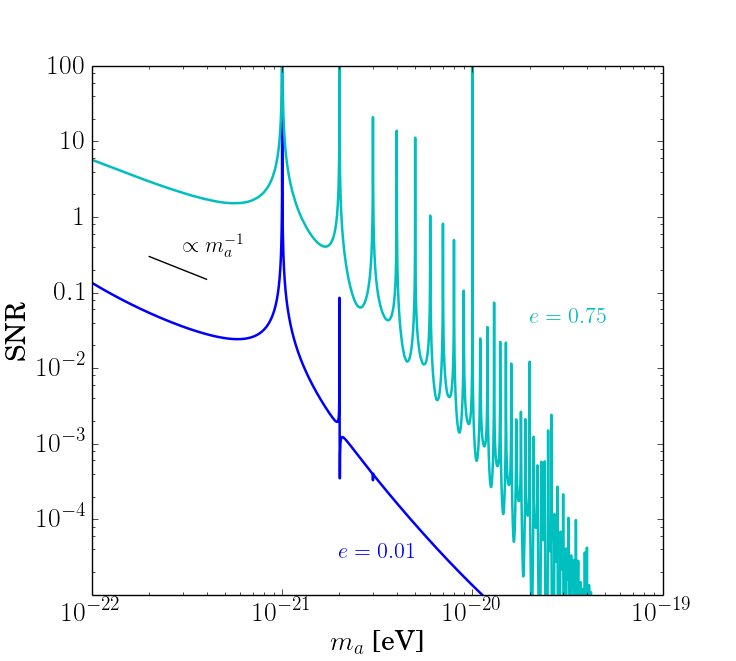}
  \caption{S/N as a function of the axion mass $m_a$ when the orbital parameters are fixed to their fiducial value (see Sec. \S\ref{sec:numerics}). Results are shown for a near-circular ($e=0.01$) and highly eccentric ($e=0.75$) orbit. For small axion masses $m_a\ll \Omega_0$, the S/N scales like $m_a^{-1}$ as indicated in the figure.}
  \label{fig:SNRma}
\end{figure}
%------------------------------------------------------------

Our prediction with the response function Eq.(\ref{eq:Responsezeroe}) is shown in Fig.\ref{fig:SNRe0} as the solid (red) curve. The overlaid dashed (magenta) curve represents the full result (i.e. including the
transient contribution). For comparison, the blue data points indicate the simulated S/N for a circular orbit. Note that we have artificially increased $\epsilon_0$ by a factor of $10^5$ in order to reduce the numerical noise.

\subsection{The case $0 < e < 1$}
\label{sec:nonzeroe}

When the eccentricity is different from zero, the perturbation to the Keplerian frequency Eq.~(\ref{eq:ChangeOrbitalFrequency}) provides the greatest contribution to the signal across the lowest-order resonances in the limit $\tobs\gg \Omega^{-1}$ since the amplitude of this effect grows like $(\Omega t)^2$.
The variation of the Keplerian frequency (unlike a simple one-dimensional harmonic oscillator for which the fundamental frequency is fixed) leads to an infinite series
of resonances located at $k\Omega= \omega_a$, with $k\in\mathbb{N}$.

\subsubsection{Perturbation to the Orbital Frequency}

The term proportional to $\mathcal{A}_n$ in Eq.(\ref{eq:ChangeOrbitalFrequency}) dominates the perturbation to the orbital frequency. Since a change in the latter affects $\theta_c$ solely, its contribution to the S/N of the Roemer time delay is given by  
\begin{widetext}
\begin{multline}
  \label{eq:leadingterm}
\frac{1}{\xobs}\int_{\xi_0}^{\xi_\text{obs}+\xi_0}\!\! d\xi\, \big(1-e\cos\xi)
\bigg[\left(\frac{\partial\tilde r}{\partial \theta_c}\right)^2 + \tilde r^2 \left(\frac{\partial\vartheta}{\partial\theta_c}\right)^2\bigg]
\times 36(\Omega t)^2\,\bigg\{\sum_{n=1}^\infty\frac{J_n(ne)}{n} \mathcal{A}_n\!\left(\frac{\Omega}{\omega_a},\alpha,\mathcal{M}_0\right)\bigg\}^2 \\
= 36\times \bigg\{\frac{1}{\xobs}\int_{\xi_0}^{\xi_\text{obs}+\xi_0}\!\!d\xi\, \big(1+e\cos\xi)\big(\mathcal{M}(\xi)-\mathcal{M}_0\big)^2\bigg\}
\times \bigg\{\sum_{n=1}^\infty\frac{J_n(ne)}{n} \mathcal{A}_n\!\left(\frac{\Omega}{\omega_a},\alpha,\mathcal{M}_0\right)\bigg\}^2\;.
\end{multline}
\end{widetext}
For large values of $\xobs\gg \xi_0$, the integral over the eccentric anomaly (as emphasized by the curly brackets) asymptotes to 
$\xi_\text{obs}^2$ in the long-time limit, with transient residuals proportional to $\xobs$ (in the best case) that can safely be neglected. Therefore, the frequency change yields a contribution
\be
\label{eq:response1}
\mathcal{R}_\infty \supset
12\, \xobs^2 \left[\sum_{n=1}^\infty \frac{J_n(ne)}{n}\mathcal{A}_n\!\left(\frac{\Omega}{\omega_a},\alpha,\mathcal{M}_0\right)\right]^2
\ee
to the response function.
As we will see shortly, it depends sensitively on the axion phase $\alpha$ and the initial condition $\mathcal{M}_0$.
Since, at fixed $\tobs$, the total observed lapse of eccentric anomaly is $\xobs \propto \Omega$, the corresponding S/N decays like $\propto a^{-1/2}$
for $a > a_0$ and eventually drops below the S/N $\propto a$ arising from the other perturbations.

\subsubsection{Instantaneous Perturbations}

To calculate the combined effect of these instantaneous perturbations, we must, here again, take into account all the resonances because they lead to the cancellation
of the zeroth-order term $\propto \Omega^0$, such that the behaviour $S/N\propto a$ is recovered in the regime $\Omega\to 0$. We demonstrate this point in Appendix \S\ref{sec:smallomegas}.
Alternatively, notice that 
\begin{widetext}
\begin{align}
  \delta q_\alpha(t) &= \frac{\partial q_\alpha}{\partial \theta_\alpha}(t) \Delta\theta_\alpha(t)+\frac{\partial q_\alpha}{\partial J_\alpha}(t)\Delta J_\alpha(t) \\
  &= \int_0^t\!dt' \, \big[q_\alpha(t),\ham_1(t')\big]_{\theta_\alpha,J_\alpha} \nonumber \\
  &= 4\pi G \rho_\text{DM}\mu \int_0^t\!dt'\,\cos(\omega_a t'+\alpha)\, r(t')\, \big[q_\alpha(t),r(t')\big]_{\theta_\alpha,J_\alpha} \nonumber \\
  &= 4\pi G\rho_\text{DM}\mu \int_{\xi_0}^{\xobs+\xi_0}\!\!\!d\xi'\,(1-e\cos\xi')\,\cos(\omega_a t'(\xi')+\alpha)\,r(\xi') \big[q_\alpha(\xi),r(\xi')\big]_{\theta_\alpha,J_\alpha} \nonumber \;.
\end{align}
\end{widetext}
where $q_\alpha=\{r,\vartheta\}$ are the polar coordinates. We have ignored the frequency change arising from $\ham_0$ (since we treat it separately). In the limit $\Omega\to 0$, the variable $t$ evolves independently of the
eccentric anomaly, which remains constant and equal to $\xi\equiv\xi_0$ during the entire observational period.
As a result, the Poisson brackets converge toward $[q_\alpha(\xi_0),r(\xi_0)]$, which must vanish by definition. Therefore, there is no contribution to the S/N proportional to $\epsilon\cdot\Omega\cdot a\sim a^{5/2}$ in the limit $\Omega\to 0$.

%------------------------------------------------------------
\begin{figure}
  \centering
  \includegraphics[width=9cm]{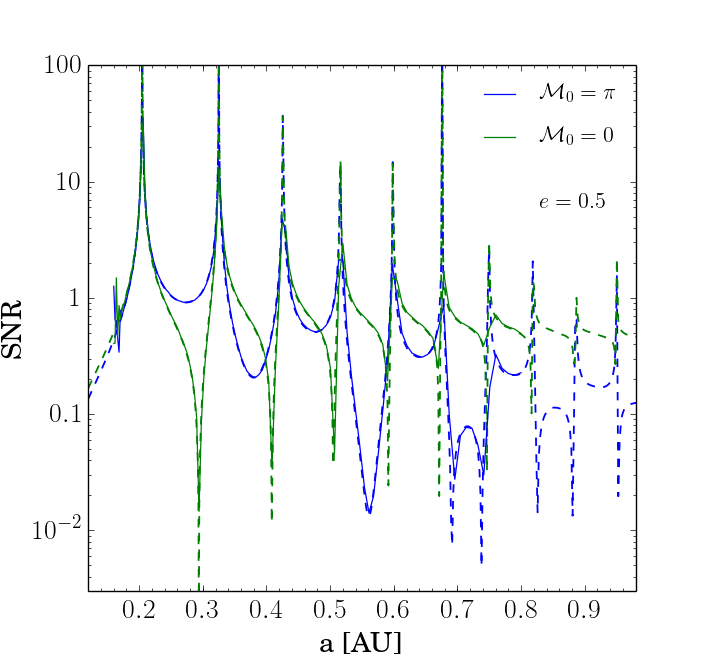}
  \caption{Dependence of the S/N on the initial conditions. The solid and dashed curves show the simulation and theoretical prediction for $\mathcal{M}_0=0$ (blue) and
    $\mathcal{M}_0=\pi$ (green). Results are shown for an unperturbed Keplerian orbit with $e=0.5$. An axion phase $\alpha=0$ is assumed throughout.}
  \label{fig:SNRics}
\end{figure}
%------------------------------------------------------------

To derive the leading nonvanishing contribution, we expand the functions $\mathcal{S}$ and $\mathcal{C}$ that appear in Eqs.(\ref{eq:DeltaTbt}) -- (\ref{eq:DeltaJct}) in the small ratio $\Omega/\omega_a$ as in Eq.(\ref{eq:SCsmallOmega}). Namely, we must retain the argument $n\Omega t$ of the trigonometric functions because $t$ can be arbitrarily large and thus $\Omega t$ is not necessarily small. 

Obtaining the exact functional dependence on $e$, $\alpha$ and $\mathcal{M}_0$ is challenging owing to the presence of a multiplicative factor of $(1-e\cos\xi)^{-1}$. A rough approximation can be derived upon treating $t$ and $\xi$ as independent variables (an approximation justified by the fact that the frequencies $\Omega$ and $\omega_a$ are vastly different) and setting $\xi=\xi_0$ (which has the advantage of removing factors of $(1-e\cos\xi)^{-1}$ in the integrand). 
Successively averaging over $t$ (with $0\leq t < 2\pi/\omega_a$) and $\xi$ (with $0\leq \xi < 2\pi/\Omega$), we eventually arrive at
\be
\label{eq:response2}
\mathcal{R}_\infty \supset 2 \left(\frac{\Omega}{\omega_a}\right)^2 \big(1+33 e^2\big)\big(2+\cos2\alpha\big) \;.
\ee
Although the dependence on $\alpha$ is certainly incorrect (in the limit $e\to 0$, any dependence on $\alpha$ should vanish as outlined in Sec. \S\ref{sec:zeroe}), this shows that the amplitude of this effect mildly increases with the eccentricity.

\subsubsection{Response Function for $0<e<1$}

In analogy with the circular case, the S/N for $0<e<1$ can be recast into the form
\be
\label{eq:SNReeorbital}
\left(\frac{S}{N}\right)^2 = \left(\frac{\epsilon\, \Omega\, a}{\omega_a}\right)^2 \frac{f_\text{obs} t_\text{obs}}{\sigma_\Delta^2c^2\Delta}\,\,
  \mathcal{R}_\infty\!\!\left(\frac{\Omega}{\omega_a},e,\alpha,\mathcal{M}_0;\Omega\tobs\right)
\ee
in which the response function $\mathcal{R}_\infty$ is the sum of Equations (\ref{eq:response1}) and (\ref{eq:response2}).
The S/N behaves like 
\be
\label{eq:SNRscaling}
\left(\frac{S}{N}\right) \propto \left\lbrace\begin{array}{ll} a^{5/2} & (a\ll a_0) \\ a^{-1/2} & (a\sim a_0)  \\ a & (a\gg a_0) \end{array}\right. \;,
\ee
where $a\sim a_0$ signifies "in the resonant region." As emphasized earlier, this behavior is consistent with the response of a harmonic oscillator to an external, oscillatory perturbation.

When the S/N is shown as a function of axion mass $m_a$ for a fixed orbital configuration (as would arise from the analysis of a given pulsar timing residuals series), Eq.(\ref{eq:SNRscaling}) gives the behavior $\propto m_a^{-1}$ for $m_a\ll \Omega_0$, and $\propto m_a^{-2}$ for $m_a\gg \Omega_0$. This behavior can be seen in Fig.\ref{fig:SNRma}, where the sensitivity curve is shown for two different eccentricities.

\subsection{Validation with Numerical Simulations}

In all subsequent illustrations, the theory curve represents the long-time asymptotic result characterized by the response function $\mathcal{R}_\infty$ (that is, we neglect the transient contribution).

Fig.\ref{fig:SNRics} demonstrates the impact of the initial mean anomaly on the S/N. When $\mathcal{M}_0=0$ or $\pi$, the response function simplifies to
\begin{align}
\mathcal{R}_\infty\!(&x,e,\alpha,\mathcal{M}_0;\xobs) =
48\, \xobs^2 \left[\sum_{n=1}^\infty\sigma(n)\frac{x J_n(ne)}{\big(1-n^2 x^2\big)}\right]^2 
\nonumber \\
&\mbox{with}\qquad \left\lbrace
\begin{array}{ll}
\sigma(n) = 1 & (\mathcal{M}_0=0) \\
\sigma(n) = (-1)^n & (\mathcal{M}_0=\pi) 
\end{array}
\right.
\end{align}
The resonant patterns vary significantly (the difference can exceed an order of magnitude at a given eccentricity) across the range of values spanned by $a$. Near the resonances, the discrepancy between the numerical data and the theoretical predictions is due to transients, which are negligible away from resonances provided that $\tobs$ is larger than a few orbital times. 

\begin{figure*}
  \centering
  \includegraphics[width=14cm]{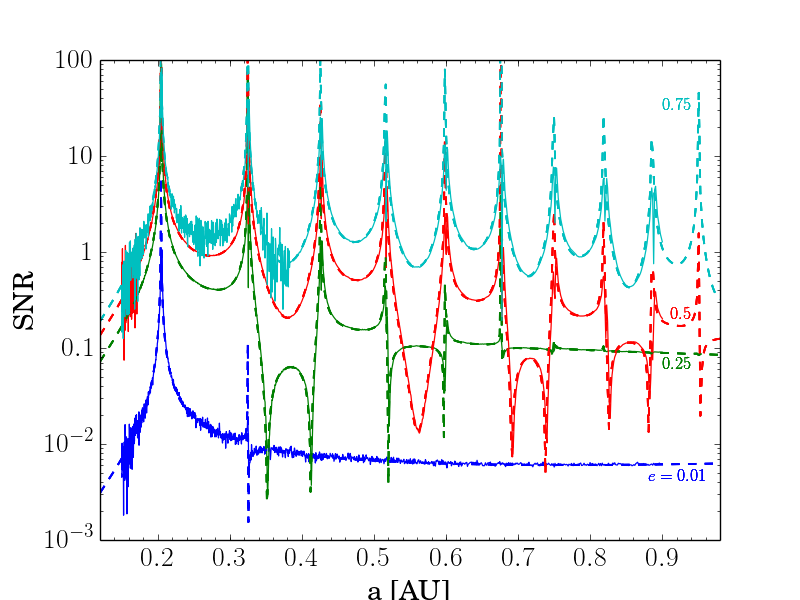}
  \caption{Signal-to-noise for various eccentricities $0<e<1$.
    The solid and dashed curves show the simulations and theoretical predictions for $e=0.01$ (blue), 0.25 (green), 0.5 (red) and 0.75 (cyan), respectively. The initial mean eccentricity is $\mathcal{M}_0=\pi$. 
    The data shown here is also displayed in Fig.4 of \cite{rozner/grishin/etal:2019}.}
  \label{fig:SNRee}
\end{figure*}

Fig.\ref{fig:SNRee} compares the prediction Eq.(\ref{eq:SNReeorbital}) (dashed curves) to the numerical result (solid curves) for different values of the
eccentricity. The model parameters are $\mathcal{M}_0=\pi$ and $\alpha=0$.
The agreement between theory and simulations is excellent for all the configurations considered here.

For the range of semi-major axis values shown here, the contribution Eq.(\ref{eq:response1}) dominates the S/N shown in Figures \ref{fig:SNRics} and \ref{fig:SNRee}. Notwithstanding, the contribution Eq.(\ref{eq:response2}) turns out to be significant for the nearly circular orbit with $e=0.01$ above $a\gtrsim 0.5$ AU.

%------------------------------------------------------------
%------------------------------------------------------------

In the classical (Newtonian) setting adopted here (see \S\ref{sec:GR} for a justification), our analytical results are valid so long as the dimensionless parameter
\begin{equation}
    \epsilon \simeq 1.43\times 10^{-15} \left(\frac{\rho_\text{\tiny DM}}{\msunpppc}\right)\left(\frac{M}{\msun}\right)^{-1}\bigg(\frac{a}{{\rm AU}}\bigg)^3 \;,
\end{equation}
is significantly smaller than unity, so that a leading-order perturbation treatment is justified. For a realistic binary pulsar mass $\sim \msun$, we have $\epsilon\lesssim 10^{-2}$ so long as the DM mass enclosed within the orbit satisfies
\begin{equation}
    \label{eq:constraint1}  
    \rho_\text{DM}a^3 \lesssim 10^{-3} \msun \;.
\end{equation}
For a semimajor axis $\lesssim 1$ AU, this holds even for DM densities orders of magnitude larger than the solar neighborhood value. Note that $\epsilon$ does not depend on either the axion mass or the orbital eccentricity.

\subsection{General Relativistic Corrections}
\label{sec:GR}

In principle, the smallness of axion perturbations to the orbital motion of a binary system requires the calculation to be carried out within a GR framework \citep[see][for recent reviews]{blanchet/etal:2014,hughes:2009}.

GR corrections to $\dot{J_c}$ enter only at 2.5PN order, but there are other corrections at lower post-Newtonian orders that induce variations in $J_a, J_b$. Overall, all these effects are calculable. Furthermore, at leading order in perturbations (relative to the Newtonian solution), most of them cancel out in equation \eqref{eq:RoemerSNRt} as the latter involves the difference between the (Newtonian or GR) solution with $\epsilon = 0$ and that with $\epsilon > 0$. Higher-order GR corrections should be negligible unless the binary is about to merge, which is not the situation considered here.

Nevertheless, the loss of energy through the emission of gravitational radiation
\citep[or the axion particles themselves if $\Omega > m_a$, see][]{poddar/etal:2019}
can affect the S/N around the resonances even when the system is far from coalescence.
Ignoring the dependence on eccentricity, the power radiated in gravitational waves during one orbital period is
\citep{peters:1964}
\be
\label{eq:peters}
P_\text{gw} = \frac{32}{5c^5}\mu^2 G^{7/3}M^{4/3} \Omega^{10/3} \;.
\ee
By comparison, the power injected by the axion coherent oscillations during one orbital period is
\be
P_a \sim \epsilon \mu \Omega^2 a^2 \cdot \Omega = \epsilon \mu G^{2/3} M^{2/3} \Omega^{5/3} \;.
\ee
This yields
\begin{align}
    \frac{P_\text{gw}}{P_a} &\simeq 4.38\times 10^{-5} \left(\frac{\rho_\text{\tiny DM}}{\msunpppc}\right)^{-1} \\
    &\qquad \times \left(\frac{\mu}{\msun}\right)\left(\frac{M}{\msun}\right)^{5/2}\bigg(\frac{a}{{\rm AU}}\bigg)^{-11/2} \nonumber \;.
\end{align}
Upon inserting our fiducial orbital parameters and assuming $\Omega\sim \omega_a$, $\mu=1\msun$, we find $P_\text{gw}/P_a\sim 3\times 10^{-4}$. This ratio is independent of the axion mass and increases with decreasing $\rho_\text{DM}$. While it is small for the large $\rho_\text{DM}$ adopted here (see Eq.(\ref{eq:epsilon0}), it would be of order unity for a dark matter density comparable to that of the solar neighborhood. In this case, we expect that the damping produced by gravitational wave emission smooths the response function around the resonances (in analogy with a simple one-dimensional damped, driven harmonic oscillator). For reasonable values of $\mu\sim M\sim \msun$, this occurs when
\begin{equation}
    \label{eq:constraint2}
    \left(\frac{\rho_\text{\tiny DM}}{\msunpppc}\right)\bigg(\frac{a}{{\rm AU}}\bigg)^{11/2}\gtrsim 10^{-5} \;.
\end{equation}
Summarizing, the calculation presented in this paper is accurate so long as the conditions (\ref{eq:constraint1}) and (\ref{eq:constraint2}) are simultaneously satisfied and, as explained in \cite{rozner/grishin/etal:2019}, the system is at least $\sqrt{\epsilon}$ away from resonances.

\subsection{Resonances}
\label{sec:resonances}

Since the energy loss $\Omega^{-1}P_\text{gw}$ is very small compared to the binding energy of the system (except for the very last stages of the merger), the orbits shrink adiabatically owing to the emission of gravitational radiation. Using the classical formula Eq.(\ref{eq:peters}), the time spent in a semi-major axis interval of width $\Delta a\sim \sqrt{\epsilon}$ is
\be
t_\text{res} \sim \frac{5}{154} \frac{c^5}{G^3\mu M^2} a^3 \Delta a \;.
\ee
Consider now the fundamental resonance centered at $\Omega=\omega_a$.
Using the techniques presented in \cite{rozner/grishin/etal:2019}, the width of the corresponding resonant region (that is, the libration region, which cannot be resolved with our perturbative approach) is
\begin{equation}
    \Delta a = \sqrt{\frac{32\epsilon J_1(e)}{3}}a_0 \;.
\end{equation}
Therefore, taking $e = 0.5$ for illustration, the time spent in the fundamental resonance is
\begin{equation}
    t_\text{res} \approx 0.05\frac{c^5a_0^4}{G^3\mu M^2}\sqrt{\epsilon_0} \;.
\end{equation}
For our fiducial parameter values, we obtain $t_\textrm{res} \approx 6.1\times 10^{7}\yr$. This shows that, for the orbital parameters adopted here, the system would stay at resonance for a duration much longer than any realistic observational time. In practice, however, the probability that a binary system will be found at resonance is very small owing to the smallness of $\sqrt{\epsilon}$. 

%%%%%%%%%%%%%%%%%%%%%%%%%%%%%%%%%%%%%%%%%%%%%%%%%%%%%%%%%%%%%%%%%%%%%
\section{Conclusions}
\label{sec:conclusions}
%%%%%%%%%%%%%%%%%%%%%%%%%%%%%%%%%%%%%%%%%%%%%%%%%%%%%%%%%%%%%%%%%%%%%

We investigated the instantaneous variations produced by the coherent oscillations of ultralight axion dark matter of mass $m_a$ on a Keplerian binary. After solving the equations of motion at first order in the (small) perturbations, we focused on the response of the binary separation to this oscillatory driving force, the amplitude of which can be constrained with pulsar timing owing to its impact on the Roemer time delay.  
The relative amplitude of this effect is proportional to $\frac{G\rho_\text{DM}}{\Omega^2}$ and thus is comparable to the relative imprint of axion oscillations on the gravitational potential \citep{khmelnitsky:2013} in the resonant region $\Omega\sim m_a$.

We computed the S/N for a measurement of instantaneous variations in the Roemer time delay, providing physical intuition whenever possible. In particular, we emphasized its similarity to the response of a harmonic oscillator to an external oscillatory driving \citep[unsurprisingly, given the duality between the Kepler problem and the two-dimensional harmonic oscillator; see][]{arnold:1990}. We outlined the dependence of such a measurement on the orbital parameters, as well as the initial axion and orbital phases. 
We compared our theoretical predictions to accurate numerical simulations and found excellent agreement for a wide range of eccentricities $0\leq e<1$. 
Although we did not consistently include the back-reaction of the binary system, which can emit energy in the form of gravitational waves, etc. \citep[see, e.g.,][for a recent discussion]{annulli/etal:2018}, we estimate for which parameter values gravitational wave emission becomes relevant.
Furthermore, we ignored the orientation of the orbital plane relative to the line of sight to the observer for simplicity, but this can be easily taken into account.

Our exact expressions furnish a useful benchmark for numerical codes and analysis procedures and, hopefully, will motivate the search for such imprints in real data. While we concentrated on dark matter in the form of a Bose-Einstein condensate of ultralight axions (for which the signal induced by oscillations in the gravitational potential is arguably small), our application of the angle-action formalism can, of course, be extended to other dark matter scenarios and/or different couplings \citep[see, e.g.,][]{blas/etal:2016,nojiri/etal:2019}. 

\acknowledgments

V.D. and Y.B.G. acknowledge support by the Israel Science Foundation (grant No. 1395/16).

\appendix

%%%%%%%%%%%%%%%%%%%%%%%%%%%%%%%%%%%%%%%%%%%%%%%%%%%%%%%%%%%%%%%%%%%%%
\section{Generating Function}
%%%%%%%%%%%%%%%%%%%%%%%%%%%%%%%%%%%%%%%%%%%%%%%%%%%%%%%%%%%%%%%%%%%%%

\label{sec:generating}

To calculate the derivatives of the polar coordinates with respect to the angle-action variables. consider the function $\gen$ generating the canonical
transformation to the Delaunay variables,
\be
\label{eq:canonicalW}
\gen\!(r,\vartheta,J_b,J_c) \equiv\int\!dr\,{\rm sgn}(\dot r)\sqrt{-\frac{\mu^2k^2}{J_c^2}+\frac{2\mu k}{r}-\frac{J_b^2}{r^2}} + J_b \vartheta \;,
\ee
and notice that the first of the two equations of canonical transformations
\be
\label{eq:canonicalEQ}
\theta_b = \frac{\partial\gen}{\partial J_b} \qquad \mbox{and}\qquad \theta_c = \frac{\partial\gen}{\partial J_c} 
\ee
involves the variables $(r,\vartheta,\theta_b,J_b,J_c)$, while the second involves only $(r,\theta_c,J_b,J_c)$.
Therefore, we can write Eq.(\ref{eq:canonicalEQ}) as $g_1=g_2=0$. The auxiliary functions $g_1$ and $g_2$ are
\begin{align}
g_1(r,\vartheta,\theta_b,J_b,J_c) &\equiv \theta_b - \frac{\partial\gen}{\partial J_b}  \;, \\
g_2(r,\theta_c,J_b,J_c) &\equiv \theta_c - \frac{\partial\gen}{\partial J_c} \nonumber \;,
\end{align}
with the understanding that all the variables should be treated as independent.
Next, we can solve $g_2=0$ for $r=r(\theta_c,J_b,J_c)$, which we subsequently substitute into $g_1=0$ to solve for
$\vartheta=\vartheta(\theta_b,\theta_c,J_b,J_c)$.

In differential form, we have
\begin{align}
  dg_1 &= \frac{\partial g_1}{\partial r}dr+\frac{\partial g_1}{\partial\vartheta}d\vartheta
  +\frac{\partial g_1}{\partial\theta_b}d\theta_b+\frac{\partial g_1}{\partial J_b}dJ_b
  +\frac{\partial g_1}{\partial J_c}dJ_c \nonumber \\
  dg_2 &= \frac{\partial g_2}{\partial r}dr
  +\frac{\partial g_2}{\partial\theta_c}d\theta_c+\frac{\partial g_2}{\partial J_b}dJ_b
  +\frac{\partial g_2}{\partial J_c}dJ_c \;.
\end{align}
Setting $dg_2=0$ implies
\begin{align}
  dr &= -\left(\frac{\partial g_2}{\partial r}\right)^{-1}\left(\frac{\partial g_2}{\partial\theta_c}d\theta_c
    + \frac{\partial g_2}{\partial J_b}dJ_b +\frac{\partial g_2}{\partial J_c}dJ_c\right)
\end{align}
Now, since Eq.(\ref{eq:canonicalEQ}) implies $\frac{\partial g_2}{\partial \theta_c}=1$, we find
\begin{equation}
\frac{\partial r}{\partial\theta_c}= - \left(\frac{\partial g_2}{\partial r}\right)^{-1}\frac{\partial g_2}{\partial\theta_c}
= - \left(\frac{\partial g_2}{\partial r}\right)^{-1}
= + \left[\frac{\partial}{\partial r}\!\!\left(\frac{\partial\gen}{\partial J_c}\right)\right]^{-1}\;.
\end{equation}
Similarly,
\begin{align}
  \frac{\partial r}{\partial J_b} &= - \left(\frac{\partial g_2}{\partial r}\right)^{-1}\frac{\partial g_2}{\partial J_b}
  = - \frac{\partial r}{\partial\theta_c}\,\frac{\partial}{\partial J_b}\!\!\left(\frac{\partial\gen}{\partial J_c}\right) \\
  \frac{\partial r}{\partial J_c} &= - \left(\frac{\partial g_2}{\partial r}\right)^{-1}\frac{\partial g_2}{\partial J_c}
  = - \frac{\partial r}{\partial\theta_c}\,\frac{\partial}{\partial J_c}\!\!\left(\frac{\partial\gen}{\partial J_c}\right) \nonumber \;.
\end{align}
To proceed further, we need
\begin{align}
  \frac{\partial\gen}{\partial J_b} &= \vartheta-{\rm sgn}(\dot r)\arccos\!\!\left(\frac{\frac{J_b^2}{\mu kr}-1}{\sqrt{1-\frac{J_b^2}{J_c^2}}}\right) \\
  \frac{\partial\gen}{\partial J_c} &= {\rm sgn}(\dot r)\Bigg[-\frac{r}{J_c}\sqrt{-\frac{\mu^2k^2}{J_c^2}+\frac{2\mu k}{r}-\frac{J_b^2}{r^2}}
  +\arccos\!\!\left(\frac{1-\frac{\mu kr}{J_c^2}}{\sqrt{1-\frac{J_b^2}{J_c^2}}}\right)\Bigg] 
  \nonumber \;.
\end{align}
In the second equality, the first term on the right-hand side vanishes at the pericenter and apocenter, i.e. $r=r_\pm=a(1\pm e)$.
Note also that $\arccos(x)$ is defined on its main branch $-\pi\leq x<\pi$. Therefore, these derivatives must be properly incremented
(that is, subtract and add Int$(\xi/2\pi+1/2)$ to the first and second line, respectively)
such that the angles $\theta_b$ and $\theta_c$ grow monotonically with time.

Parameterizing the unperturbed trajectory with the eccentric anomaly $\xi$, the radial coordinate reads $r(\xi)=a(1-e\cos\xi)$ and the partial derivatives of $r$ reduce to
\begin{align}
  \label{eq:dradial}
  \frac{\partial r}{\partial\theta_b}\bigg\lvert_0 &= 0 \\
  \frac{\partial r}{\partial\theta_c}\bigg\lvert_0 &= \frac{ae\sin\xi}{\big(1-e\cos\xi\big)} \nonumber \\
  \frac{\partial r}{\partial J_b}\bigg\lvert_0 &= \frac{a}{J_c}\frac{\sqrt{1-e^2}}{e}\frac{\big(e-\cos\xi\big)}{\big(e\cos\xi-1\big)}\nonumber \\
  \frac{\partial r}{\partial J_c}\bigg\lvert_0 &= \frac{a}{J_c} \frac{\big(-3e+\cos\xi+3e^2\cos\xi-e^3\cos(2\xi)\big)}{e\big(e\cos\xi-1\big)}
  \nonumber \;.
\end{align}
In this derivation, it is essential to take into account the multiplicative factor of sgn$(\dot r)$=sgn($\sin\xi$) in the generating function
$\gen(r,\vartheta,J_b,J_c)$ as it ensures that all the partial derivatives are continuous functions of $\xi$.

The calculation of the derivatives $\partial\vartheta/\partial\theta_\alpha$ and $\partial\vartheta/\partial J_\alpha$ proceeds analogously. Setting
$dg_1=0$, substituting $r=r(\theta_c,J_b,J_c)$ and taking advantage of the fact that
$\frac{\partial g_1}{\partial\vartheta}= -\frac{\partial}{\partial\vartheta}\frac{\partial\gen}{\partial J_b}=-1$, we obtain
\begin{equation}
d\vartheta = \bigg[\frac{\partial g_1}{\partial\theta_b}d\theta_b+\frac{\partial g_1}{\partial r}\, \frac{\partial r}{\partial\theta_c}d\theta_c
  + \left(\frac{\partial g_1}{\partial J_b}+\frac{\partial g_1}{\partial r}\frac{\partial r}{\partial J_b}\right) dJ_b +\left(\frac{\partial g_1}{\partial J_c}+\frac{\partial g_1}{\partial r}\frac{\partial r}{\partial J_c}\right) dJ_c\bigg] \;.
\end{equation}
For instance, we read off
\begin{equation}
  \frac{\partial \vartheta}{\partial J_b} = \frac{\partial g_1}{\partial J_b}+\frac{\partial g_1}{\partial r}\frac{\partial r}{\partial J_b}
  = -\frac{\partial}{\partial J_b}\!\!\left(\frac{\partial\gen}{\partial J_b}\right) - \frac{\partial r}{\partial J_b}\,
  \frac{\partial}{\partial r}\!\!\left(\frac{\partial\gen}{\partial J_b}\right) \;.
\end{equation}
After some algebra, we arrive at
\begin{align}
  \label{eq:dtheta}
  \frac{\partial\vartheta}{\partial\theta_b}\bigg\lvert_0 &= 1 \\
  \frac{\partial\vartheta}{\partial\theta_c}\bigg\lvert_0 &= \frac{\sqrt{1-e^2}}{\big(1-e\cos\xi\big)^2} \nonumber \\
  \frac{\partial\vartheta}{\partial J_b}\bigg\lvert_0 &= \frac{1}{J_c}\frac{\big(-2+e^2+e\cos\xi\big)}{e\big(1-e\cos\xi\big)^2}\sin\xi
  \nonumber \\
  \frac{\partial\vartheta}{\partial J_c}\bigg\lvert_0
  &= \frac{1}{J_c}\frac{\sqrt{1-e^2}}{e}\frac{\big(2-e^2-e\cos\xi\big)}{\big(1-e\cos\xi\big)^2}\sin\xi \nonumber \;.
\end{align}
One can check that the following (equal-time) Poisson bracket vanishes identically for any $\xi$,
\be
\big[r,\vartheta\big]_{\theta_\alpha,J_\alpha} \equiv 0 \;.
\ee
This indicates that the various partial derivatives we computed are consistent with a canonical transformation, as it should be. Note that $\big[r,\vartheta\big]$ does not generally vanish when $r$ and $\vartheta$ are evaluated at different times on the physical trajectory.

%%%%%%%%%%%%%%%%%%%%%%%%%%%%%%%%%%%%%%%%%%%%%%%%%%%%%%%%%%%%%%%%%%%%%
\section{Development at Small Eccentricities}
%%%%%%%%%%%%%%%%%%%%%%%%%%%%%%%%%%%%%%%%%%%%%%%%%%%%%%%%%%%%%%%%%%%%%

\label{sec:smalleccentricities}

To proceed, we specialize the partial derivatives of the polar coordinates $(r,\vartheta)$ with respect to the angles and actions (along the unperturbed trajectory, which is now a circular orbit) to the case $e\to 0$:
\begin{align}
  \frac{\partial r}{\partial\theta_b}\bigg\lvert_0 &= 0 \\
  \frac{\partial r}{\partial\theta_c}\bigg\lvert_0 &\approx ae\sin\xi + ae^2\cos\xi\sin\xi %+\mathcal{O}(e^3) 
  \nonumber \\
  \frac{\partial r}{\partial J_b}\bigg\lvert_0 &\approx \frac{a}{e J_c}\cos\xi -\frac{a}{J_c}\sin^2\xi
  +\frac{a e }{2J_c}\big(\cos 2\xi- 2\big)\cos\xi %+ \mathcal{O}(e^2) 
  \nonumber \\
  \frac{\partial r}{\partial J_c}\bigg\lvert_0 &\approx -\frac{a}{e J_c}\cos\xi
  +\frac{a}{2J_c}\big(5-\cos 2\xi\big)-\frac{ae}{J_c}\cos^3\xi %+ \mathcal{O}(e^2) 
  \nonumber \\
  \frac{\partial \vartheta}{\partial\theta_b}\bigg\lvert_0 &= 1 \nonumber \\
  \frac{\partial \vartheta}{\partial\theta_c}\bigg\lvert_0 &\approx 1+2e\cos\xi -\left(\frac{1}{2}-3\cos^2\xi\right)e^2 %+\mathcal{O}(e^3)
  \nonumber \\
  \frac{\partial \vartheta}{\partial J_b}\bigg\lvert_0 &\approx -\frac{2}{eJ_c}\sin\xi-\frac{3}{J_c}\sin\xi\,\cos\xi -\frac{e}{J_c}\big(1+2\cos 2\xi\big)\sin\xi 
  %+ \mathcal{O}(e^2) 
  \nonumber \\
  \frac{\partial \vartheta}{\partial J_c}\bigg\lvert_0 &\approx \frac{2}{eJ_c}\sin\xi+\frac{3}{J_c}\sin\xi\,\cos\xi
  +\frac{e}{J_c}\big(\sin 3\xi-\sin\xi\big) %+ \mathcal{O}(e^2) 
  \nonumber \;.
\end{align}
In each expression, we retained terms up to order $e$ except for the partial derivatives relative to $\theta_c$, for which we include terms up to order $e^2$ (because the
perturbation $\Delta\theta_c$ features a contribution of order $e^{-1}$). 
Although the derivatives relative to the actions diverge in the limit $e\to 0$, the relation $[r,\vartheta]_{\theta_\alpha,J_\alpha}=0$ is always satisfied along the physical trajectory (i.e. when $r$ and $\vartheta$ are evaluated at a fixed $\xi$).
We now turn to the expressions for $\Delta\theta_\alpha$ and $\Delta J_\alpha$, and use the fact that, for small $ne\ll 1$, the Bessel functions behave like
$J_n(ne)\sim (ne)^n$. In particular, $J_1(e)\approx e/2$.
Writing $J_n'(n e) = J_{n-1}(ne)-\frac{1}{e}J_n(ne)$, Taylor-expanding the Bessel functions in the small argument limit, and retaining terms up to order $e$
(since the partial derivatives of $r$ and $\vartheta$ with respect to the angles $\theta_b$ and $\theta_c$ are at best of order $e^0$), we find
%\begin{widetext}
\begin{align}
  \Delta\theta_b &\stackrel{e\to 0}{=}
  \epsilon \left(\frac{\Omega}{\omega_a}\right) \bigg\{ \frac{\mathcal{S}_{11}^{(+)}}{e}+ \frac{\mathcal{S}_{21}^{(+)}}{2}
  -3\Big(\sin(\omega_at+\alpha)-\sin\alpha\Big)+ \frac{1}{8}\Big(-7\mathcal{S}_{11}^{(+)}+3\mathcal{S}_{31}^{(+)}\Big) e
  + \mathcal{O}(e^2) \bigg\} \\
  \Delta\theta_c &\stackrel{e\to 0}{=}
  -\epsilon\left(\frac{\Omega}{\omega_a}\right)\bigg\{\frac{\mathcal{S}_{11}^{(+)}}{e}+\frac{\mathcal{S}_{21}^{(+)}}{2}
  -7\Big(\sin(\omega_at+\alpha)-\sin\alpha\Big) + \bigg[\frac{3}{8}\Big(7\mathcal{S}_{11}^{(+)}+\mathcal{S}_{31}^{(+)}\Big)
    +3 \left(\frac{\Omega}{\omega_a}\right) \mathcal{S}_{12}^{(-)}-3\Omega t\,\mathcal{A}_1
    \bigg] e + \mathcal{O}(e^2) \bigg\} \nonumber \;.
\end{align}
All the functions $\mathcal{S}_{1q}^{(\pm)}$ and $\mathcal{C}_{1q}^{(\pm)}$ that appear in the previous expressions are evaluated at $t>0$.
Furthermore, although the individual deviations $\Delta\theta_b$ and $\Delta\theta_c$ diverge in the limit $e\to 0$, their contribution to the displacement
$\delta\vr$ is always well behaved since the latter depends on
\begin{align}
  \frac{\partial \vartheta}{\partial \theta_b}\Delta\theta_b + \frac{\partial\vartheta}{\partial\theta_c}\Delta\theta_c &=
  \epsilon\left(\frac{\Omega}{\omega_a}\right)\bigg\{4\Big(\sin(\omega_at+\alpha)-\sin\alpha\Big)-2 \cos\xi\, \mathcal{S}_{11}^{(+)} +
  \bigg[- \Big(3\mathcal{S}_{11}^{(+)}+3\cos^2\!\xi\,\mathcal{S}_{11}^{(+)}+\cos\xi\,\mathcal{S}_{21}^{(+)}\Big) \nonumber \\
    &\quad + 14\Big(\sin(\omega_at+\alpha)-\sin\alpha\Big)
    -3\left(\frac{\Omega}{\omega_a}\right)\mathcal{S}_{12}^{(-)}+3\Omega t\,\mathcal{A}_1\bigg] e
    +\mathcal{O}(e^2)\bigg\}\;.
\end{align}
For the action variables, we have $\Delta J_b\equiv 0$ while
\begin{equation}
  \Delta J_c \stackrel{e\to 0}{=} \epsilon J_c\left(\frac{\Omega}{\omega_a}\right)\bigg[\mathcal{C}_{11}^{(-)}e
  + \frac{1}{2}\mathcal{C}_{21}^{(-)} e^2 + \mathcal{O}(e^3)\bigg] \;.
\end{equation}
The term linear in $e$ in $\Delta J_c$ combines with that proportional to $e^{-1}$ in $\frac{\partial\theta_c}{\partial J_c}$ to give an $e^0$ contribution. More precisely, 
\be
  \frac{\partial \vartheta}{\partial J_b}\Delta J_b  + \frac{\partial\vartheta}{\partial J_c}\Delta J_c =
  \epsilon\left(\frac{\Omega}{\omega_a}\right)\bigg[2\sin\xi\,\mathcal{C}_{11}^{(-)} + \sin\xi\Big(3\cos\xi\,\mathcal{C}_{11}^{(-)}+\mathcal{C}_{21}^{(-)}\Big) e + \mathcal{O}(e^2) \bigg] \;.
\ee
Applying the same analysis to the radial coordinate $r$ eventually leads to
\begin{align}
  \frac{\partial r}{\partial \theta_b}\Delta\theta_b + \frac{\partial r}{\partial\theta_c}\Delta\theta_c &=
  - \epsilon a\left(\frac{\Omega}{\omega_a}\right) \bigg\{\sin\xi\,\mathcal{S}_{11}^{(+)} +\sin\xi\bigg[\cos\xi\,\mathcal{S}_{11}^{(+)}+\frac{1}{2}\mathcal{S}_{21}^{(+)}-3\Big(\sin(\omega_at+\alpha)-\sin\alpha\Big)\bigg]e+\mathcal{O}(e^2)\bigg\}
  \nonumber \\
  \frac{\partial r}{\partial J_b}\Delta J_b + \frac{\partial r}{\partial J_c}\Delta J_c &=
  -\epsilon a \left(\frac{\Omega}{\omega_a}\right) \bigg\{\cos\xi\,\mathcal{C}_{11}^{(-)}
  +\left[\frac{1}{2}\Big(\cos\xi\mathcal{C}_{21}^{(-)}+\big(\cos2\xi-3\big)\mathcal{C}_{11}^{(-)}\Big)
    -\mathcal{C}_{11}^{(-)}\right]e +\mathcal{O}(e^2)\bigg\} \;.
\end{align}
%\end{widetext}
Putting all this together, the square $|\delta\tilde\vr|^2$ of the normalized perturbed displacement can be recast in the form of Eq.~(\ref{eq:tildedeltare0}).

%%%%%%%%%%%%%%%%%%%%%%%%%%%%%%%%%%%%%%%%%%%%%%%%%%%%%%%%%%%%%%%%%%%%%
\section{Perturbations in the Limit $\Omega\to 0$}
%%%%%%%%%%%%%%%%%%%%%%%%%%%%%%%%%%%%%%%%%%%%%%%%%%%%%%%%%%%%%%%%%%%%%

\label{sec:smallomegas}

We consider the regime $\Omega\ll\omega_a$ and expand the functions $\mathcal{S}_{n1}^{(+)}$ and $\mathcal{C}_{n1}^{(-)}$ that appear in Eqs.(\ref{eq:DeltaTbt}) -- (\ref{eq:DeltaJct}) accordingly to obtain
\begin{align}
\label{eq:SCsmallOmega}
\mathcal{S}_{n1}^{(+)} &\approx 2 \Big(\sin(\omega_a t+\alpha)\cos(n\mathcal{M})-\sin \alpha\cos(n\mathcal{M}_0)\Big) - 2 n \left(\frac{\Omega}{\omega_a}\right) \Big(\cos(\omega_a t+\alpha)\sin(n\mathcal{M})-\cos\alpha\sin(n\mathcal{M}_0)\Big) \\
\mathcal{C}_{n1}^{(-)} &\approx -2 \Big(\sin(\omega_a t+\alpha)\sin(n\mathcal{M})-\sin \alpha\sin(n\mathcal{M}_0) \Big)-2 n \left(\frac{\Omega}{\omega_a}\right)\Big(\cos(\omega_a t+\alpha)\cos(n\mathcal{M})-\cos\alpha\cos(n\mathcal{M}_0) \Big)\nonumber \;.
\end{align}
In the limit $\Omega\to 0$, only the first term subsists on the right-hand side of Eq.(\ref{eq:SCsmallOmega}). Furthermore, $\mathcal{M}\approx \mathcal{M}_0$ since the mean anomaly does not change appreciably during the observational period.
Therefore, we find
\begin{align}
\label{eq:SCsmallOmega}
\mathcal{S}_{n1}^{(+)} &\stackrel{\Omega\to 0}{=} 2 \Big(\sin(\omega_a t+\alpha)-\sin \alpha\Big)\cos(n\mathcal{M}_0)\\
\mathcal{C}_{n1}^{(-)} &\stackrel{\Omega\to 0}{=} -2 \Big(\sin(\omega_a t+\alpha)-\sin \alpha\Big)\sin(n\mathcal{M}_0) \nonumber \;.
\end{align}
Substituting these expressions into Eqs.(\ref{eq:DeltaTbt}) -- (\ref{eq:DeltaJct}) and taking advantage of the relations
\begin{align}
  \sum_{n=1}^\infty J_n(ne)\cos(n\mathcal{M}) &= \frac{e\cos\xi}{2\big(1-e\cos\xi\big)} \\
  \sum_{n=1}^\infty \frac{J_n(ne)}{n}\sin(n\mathcal{M}) &= \frac{e}{2}\sin\xi \nonumber \;,
\end{align}
the series expansions of $\Delta\theta_b$, $\Delta\theta_c$ and $\Delta J_c$ become
\begin{align}
  \label{eq:DeltaTJsmallOmega}
  \Delta\theta_b &\approx 2 \epsilon \left(\frac{\Omega}{\omega_a}\right) \frac{\sqrt{1-e^2}}{e}\big(\cos\xi_0-e\big) \Big(\sin(\omega_a t+\alpha)-\sin\alpha\Big) + \dots \\
  \Delta\theta_c &\approx
  2\epsilon \left(\frac{\Omega}{\omega_a}\right) \left[3-\left(\frac{1}{e}+3e\right)\cos\xi_0+e^2\cos(2\xi_0)\right]\Big(\sin(\omega_a t+\alpha)-\sin\alpha\Big) + \dots \nonumber \\
  \Delta J_c &\approx -2\epsilon J_c\left(\frac{\Omega}{\omega_a}\right)e\sin\xi_0 \Big(\sin(\omega_a t+\alpha)-\sin\alpha\Big) + \dots \nonumber
\end{align}
in the limit $\Omega\to 0$. On setting $\xi=\xi_0$ in Eqs.~(\ref{eq:dradial}) and (\ref{eq:dtheta}) (but allowing $t$ to grow freely) and taking into account the leading contribution in Eq.(\ref{eq:DeltaTJsmallOmega}) solely, we can check that
\begin{align}
  \delta r &= \frac{\partial r}{\partial\theta_c}(\xi_0)\Delta\theta_c + \frac{\partial r}{\partial J_c}(\xi_0) \Delta J_c \\
  &= \epsilon a\left(\frac{\Omega}{\omega_a}\right)\bigg\{\left(\frac{e\sin\xi_0}{1-e\cos\xi_0}\right)2\left[3-\left(\frac{1}{3}+3e\right)\cos\xi_0+e^2\cos(2\xi_0)\right]
  \nonumber \\
  &\qquad - \frac{\big(-3e+\cos\xi_0+3e^2\cos\xi_0-e^3\cos(2\xi_0)\big)}{e\big(e\cos\xi_0-1\big)}2 e\sin\xi_0\bigg\} \Big(\sin(\omega_a t+\alpha)-\sin\alpha\Big) \nonumber \\
  &\equiv 0 \nonumber \;,
\end{align}
as the term in curly brackets identically vanishes regardless the value of $\xi_0$.
Similar manipulations lead to $\delta\vartheta\equiv 0$ under the same assumptions. 
This implies that the response function must scale like $\mathcal{R}_\infty\propto (\Omega/\omega_a)^2$ for small $\Omega/\omega_a$.

\bibliography{references}

\end{document}